\documentstyle[12pt]{article}
\def\vx{\vec{x}}
\def\vp{\vec{p}}
\def\vq{\vec{q}}
\def\vr{\vec{r}}
\def\haf{\textstyle{1\over2}}

\newcommand{\boldpi}{\mbox{\boldmath $\pi$}}
\newcommand{\boldtau}{\mbox{\boldmath $\tau$}}
\newcommand{\vecsig}{\vec{\sigma}}
\newcommand{\fpi}{f_{\pi}}
\newcommand{\mpi}{m_{\pi}}

\newskip\humongous \humongous=0pt plus 1000pt minus 1000pt
\def\caja{\mathsurround=0pt}

\newif\ifdtup
\def\panorama{\global\dtuptrue \openup1\jot \caja
        \everycr{\noalign{\ifdtup \global\dtupfalse
        \vskip-\lineskiplimit \vskip\normallineskiplimit
        \else \penalty\interdisplaylinepenalty \fi}}}
\def\eqalignno#1{\panorama \tabskip=\humongous
        \halign to\displaywidth{\hfil$\displaystyle{##}$
        \tabskip=0pt&$\displaystyle{{}##}$\hfil
        \tabskip=\humongous&\llap{$##$}\tabskip=0pt
        \crcr#1\crcr}}
\begin{document}
\vspace*{-1.0in}
\hspace{\fill} \fbox{\bf LA-UR-99-296-REV}

\begin{center}

{\Large {\bf Equivalence of Nonstatic Two-Pion-Exchange Nucleon-Nucleon 
Potentials}}\\

\vspace*{0.80in}

J. L.\ Friar \\
Theoretical Division \\
Los Alamos National Laboratory \\
Los Alamos, NM  87545 USA\\

\vspace*{0.15in}
\end{center}

\vspace*{0.50in}

\begin{abstract}
Off-shell aspects of the one-pion-exchange potential and their relationship to
different forms of the nonstatic (subleading-order) chiral two-pion-exchange 
nucleon-nucleon potential are discussed.  Various types of off-shell behavior 
are categorized and numerous examples are given.  Recently derived potentials 
based on chiral approaches are supplemented by a rather general form of the
two-pion-exchange potential that is derived using old-fashioned methods and
exhibits off-shell dependence.  The latter potential is closely related to a 
general form of one-pion-exchange relativistic corrections and nonstatic 
two-pion-exchange three-nucleon forces developed long ago.

\end{abstract}
\pagebreak
\section{Introduction}
Off-shell effects in nuclear potentials have long posed conceptual and technical
problems in nuclear physics\cite{tm,3nf}.  Although the off-shell problem often
is caused by poorly posed input (and is therefore largely insoluble), even if
the input is completely specified (e.g., a Lagrangian) there can still be
ambiguity in final forms obtained for potentials.  We will focus on the 
ambiguities, which are basically {\bf choices} made by theorists. Thus, many
different potentials are correct (to some order), but have different forms (in
part). Nevertheless, these different forms lead to {\bf equivalent} observables.
We refer collectively to these ambiguities as ``off-shell'' effects. We wish to 
emphasize that different formalisms make these choices automatically. It is only
in trying to relate different calculations that these ambiguities must be 
understood. That is our primary purpose in this work.

There are three basic types of ambiguity that have arisen in calculating nuclear
potentials\cite{quasi}.  The first type is caused by an energy-dependent
potential, $V(E)$, where $E$ is the eigenvalue of the appropriate equation of
which $V(E)$ is part. Such forms occur naturally\cite{mec,cf} when expanding
energy denominators that occur in Schr\"{o}dinger perturbation-theory treatments
of mesons that are exchanged between nucleons: $V(E) \sim \frac{1}{E_m} J
\frac{1}{E_m + (H - E)}\,J$, where $J$ is a meson-nucleon vertex (the two $J$'s
refer to different nucleons), $E_m$ is the transiting-meson energy and $H-E$ is
the nucleons' energy difference between intermediate and initial states. 
Expanding in powers of $(H-E)/E_m$, one finds the usual static potential
$(J^2/E^2_m)$ plus energy-dependent corrections. This is perfectly satisfactory
(the Sturm-Liouville equation\cite{sl} is of this type), but energy-dependence 
is difficult to implement in anything other than the two-nucleon system.  This
energy-dependence was the origin of the Brueckner-Watson\cite{bw} vs.
TMO\cite{tmo} controversy long ago (reviewed in \cite{fc}), and many results
from chiral perturbation theory have this form\cite{texas}.  The cost of this
dependence is sufficiently high (particularly for $A \geq 3$) that we do not
recommend keeping it, but rather one should eliminate it by any of various means
developed over the years\cite{quasi}. This has been done in all of the results 
of Section 2. We will not further consider this type of off-shell dependence.

The second type of off-shell dependence arises from unitary ambiguities in the
form of the potential, often from different choices of field variables\cite{cf,
3nf} in the {\bf same} Lagrangian.  Thus if $H$ is the nuclear Hamiltonian
$$
(E - H) \Psi = 0 \, , \eqno(1)
$$
a transformation $\Psi^\prime = e^{i U} \Psi \cong (1 + i U)\Psi$ (for small 
$U$) produces 
$$
(E-H^\prime)\Psi^\prime = 0 \, , \eqno(2a)
$$
where the transformed Hamiltonian
$$
H^\prime \cong H - i[H,U] \, \eqno(2b)
$$
is a satisfactory variant of $H$ (for small $U$). Such transformations effect
gauge transformations in electromagnetic problems and occur naturally in
relativistic corrections to nuclear potentials\cite{s-o,mec,cf}.

One specific example of the latter is the quasipotential ambiguity\cite{quasi},
determined by an off-shell parameter $\nu$.  When a meson of mass $m$ is
exchanged between two nucleons (``$1$'' and ``$2$'') in an arbitrary reference
frame, the relativistic propagator has the form
$$
\frac{1}{q_0^2 - (\vq^{\; 2} + m^2)} \cong 
- \frac{1}{\vq^{\; 2} + m^2} - \frac{q^2_0}{(\vq^{\; 2} + m^2)^2} \ , \eqno(3)
$$
where $q_0^2$ is nonvanishing in general and is given by $\Delta E_1^2$ or 
$\Delta E^2_2$ or $-\Delta E_1\Delta E_2$, where $\Delta E_i$ is the energy 
transferred by the meson to nucleon ``$i$''.  Each choice leads to a different 
off-shell potential, corresponding to the different ways of projecting out 
$q_0$, the time component of the four-momentum vector. A general result for 
$q_0^2$ is given by the linear combination: [$(\Delta E_1 + \Delta E_2)^2 (1 -
\nu) -2 \Delta E_1 \Delta E_2$], which specifies the parameter 
$\nu$ \cite{quasi} and demonstrates that the effect is a unitary ambiguity. 
Common choices of the parameter $\nu$ are $\nu = 0$ (standard)\cite{quasi,s-o}, 
$\nu = 1/2$ (no retardation)\cite{mec,rel}, and $\nu = 1$ (soft). Most 
techniques lead to $\nu = 0$, although a Foldy-Wouthuysen procedure\cite{min} 
leads to $\nu = 1$. The popular $\nu=1/2$ choice\cite{cf} simplifies the nuclear
potential. Note that $\Delta E^2_i$ is of order $(1/M^2)$, where $M$ is the 
nucleon mass. This generates a relativistic correction to the potential.

Another unitary ambiguity arises when using pseudoscalar (PS) or pseudovector
(PV) relativistic pion-nucleon interactions.  The on-shell forms are identical,
but they differ off-shell.  These differences can be subsumed\cite{mec,cf} to
order $(1/M^2)$ by a collection of off-shell operators proportional to a
parameter $\mu$. Common choices are $\mu = -1$ (PS free spinors)\cite{tm,s-o,
bonn}, $\mu = 0$ (minimal nonlocality)\cite{atom,jun}, $\mu = 1$ (PV)\cite{s-o},
and $\mu = 3$ (soft)\cite{mec}. Conventional Chiral Perturbation Theory 
(CPT)\cite{cpt,bkm} expansions (term-by-term chirality) correspond to $\mu=1$
and are preferred for that reason, although other criteria having to do with the
final form of OPEP may mandate other values (viz., $\mu=0$, discussed below).

The third type of ambiguity is the form ambiguity.  Potentials only make sense
in the context of a scheme for iterating them to produce a final (to all orders)
result for a binding energy or scattering amplitude.  The nonrelativistic
paradigm is to add the kinetic energy $T_{NR} = \vp^{\; 2}/2 M$ to the potential
$V$ to obtain the Hamiltonian $H_{NR}$ that is used in Eq.~(1) to obtain the 
desired observable. The naive (but obvious) relativistic generalization of this 
is to form the free energy\cite{mec}
$$
T = \sum^A_{i=1} \sqrt{\vp_i^{\; 2} + M^2} \rightarrow 2 \sqrt{\vp^{\; 2} + M^2}
\, ,  \eqno(4)
$$
for a collection of identical-mass nucleons, where the second form applies only
to the two-nucleon problem in its center-of-mass frame.  Then $H_R = T + V$ is
the appropriate relativistic form of the energy to be used in Eq.~(1).  A
one-time formalism (linear in $E$) is always possible if one freezes out
antinucleon degrees of freedom or by using a variety of projection schemes,
which typically lead to different quasipotential equations\cite{quasi}.

An example of a quasipotential method is the popular
Blankenbecler-Sugar\cite{bs} (BS) procedure, which is based on a two-nucleon
Green's function of nonrelativistic form. Although this is a well-developed 
procedure that casts field-theory calculations into a Schr\"odinger-like 
appearance, the relationship between BS results and more conventional ones 
(using Eq.~(4) for $T$) can be obtained by means of a trick\cite{fritz}.The 
two-body relativistic Schr\"{o}dinger equation (RSE; for lack of a better name) 
corresponding to the second form in Eq.~(4) is:
$$
\left[ 2\sqrt{\vp^{\; 2} + M^2} + V(\vr)  \ \right] \Psi_R = 2\sqrt{k^2 + M^2}
\ \Psi_R  \, , \eqno(5)
$$
where $k$ is the momentum corresponding to the energy eigenvalue $E$ and $\vr$ 
is the nucleons' separation.  Squaring both sides, subtracting $4 M^2$ from both
sides, and dividing by 4M lead to the remarkable and 
{\bf equivalent}\cite{fritz}
$$
\left [\frac{\vp^{\; 2}}{M} + \overline{V} (\vr) \right ] \Psi_{BS} = 
\frac{k^2}{M} \Psi_{BS} \, ,  \eqno (6)
$$
where
$$
\overline{V}(\vr) = \left\{ \frac{\sqrt{\vp^{\; 2} + M^2}}{2M} , V(\vr) \right\}
+ \frac{V^2}{4M}  \eqno(7)
$$
is the effective potential to be used in what appears to be a nonrelativistic
Schr\"{o}dinger equation.  This equation requires relativistic kinematics (e.g.,
in relating $k$ and $E$), however, and is not nonrelativistic.  This procedure 
(using Eq.~(6)) is at the heart of the Nijmegen Partial-Wave-Analysis (PWA) 
program\cite{psa}for treating nucleon-nucleon scattering.

Any form change in the equation redefines the effective potential, which is
hardly a surprise.  Note the $V^2/4M$ term, which will be seen again since it
typically sets the scale for corrections resulting from changes of form, and the
factor $\sim (\frac{E}{M})$ that multiplies $V$.  Such factors occur
everywhere when calculating potentials.  This factor repositions relativistic
corrections from $T$ to $\overline{V}$.  Clearly other form changes are
possible.  Although Eq.~(6) is a decided advantage in treating the two-nucleon
problem, it is not of any obvious use for the many-nucleon problem (there will
be cross terms between the kinetic energies of different nucleons).  A recent
major advance has been the ability to handle $\sqrt{p^2 + m^2}$ in configuration
space\cite{rll}, since modeling momentum-dependent operators in that space is
the single most challenging numerical problem.

In Section 2 we review selected previous calculations and derive a new general 
form of the nonstatic (subleading order in CPT) two-pion-exchange 
nucleon-nucleon potential that manifests the unitary ambiguities discussed 
above. This form subsumes several previously calculated potentials as well as
a number of cases not previously considered, and allows all of these results to 
be compared. In Section 3 results for specific off-shell choices are discussed, 
including the important case of the Nijmegen PWA program. Our summary is
presented in Section 4.

\section{Calculation}

It should be obvious from the previous examples that we have been discussing
variants of relativistic corrections.  Using either the old rules of 
scale\cite{cf} (counting $T \sim 1/M$ or $1/\Lambda$, where $\Lambda$ is the
large-mass scale of QCD, and $V \sim T \sim 1/M$) or more general and
sophisticated power-counting rules\cite{cpt,ndpc} leads to the same conclusion. 
Relativistic corrections to the nucleon kinetic energy $(\sim 1/M^3$ or 
$1/\Lambda^3)$ can be juggled into the BS potential $(\sim V/M^2 \sim 
1/\Lambda^3)$ or the BS correction term $(V^2/4M \sim 1/\Lambda^3)$.  The 
retardation corrections have the same intrinsic size $(V/M^2 \sim 1/\Lambda^3)$.
This defines the limits of the expansions (in $1/M$) that will be made below.

With the exception of recent CPT calculations (or other chiral variants) much of
the work on one- and two-pion-exchange potentials is quite old\cite{tom}. In
particular, the work of TMO\cite{tmo} and Sugawara and Okubo (S-O)\cite{s-o}
stand out in their technical clarity.  Both developed an energy-independent
potential.  The static (i.e., leading order in CPT) two-pion-exchange potential 
$(V^0_{2\pi})$ was developed
by TMO and was recently reviewed in Ref.\cite{fc} in the context of
energy-dependent alternatives.  The leading-order nonstatic (i.e., subleading 
order in CPT) corrections $(\Delta
V_{2\pi})$ were developed in S-O, even though they had no credible theory of the
interactions of pions and nucleons, which came later.  Nevertheless, with the
exception of missing seagull terms ($ \pi \pi$-$N$ interactions) one of their
models (PV coupling) produced correct and complete results.  At the time of S-O
it was known that nucleon-antinucleon ``pair'' terms in PS coupling were
unphysically large and had to be suppressed (they were simply thrown away). The
reason for this suppression is now known to be chiral symmetry, which explains
the success of their PV coupling model (derivative coupling with negligible pair
terms). In addition S-O supplemented their PV Lagrangian with two seagull
interactions, one with a single derivative of Weinberg-Tomozawa type\cite{wt} 
and another (with no derivatives) of $\sigma$-term type ($c_1$ in Eq.~(9)
below). Their results were correct for these interactions, as well. They did not
consider seagulls with two derivatives, such as the $c_3$ and $c_4$ terms in
Eq.~(9).

The work of S-O also emphasized (by example) that no unique form for $\Delta
V_{2\pi}$ exists without {\bf choosing} an appropriate off-shell form for
$\Delta V_\pi$, the relativistic corrections to the one-pion-exchange potential
$V^0_\pi$ (OPEP).  Moreover, this choice also affects the nonstatic
two-pion-exchange three-nucleon forces (of nominal size $V^{0\; 2}_\pi/M$).  The
latter was developed in Ref.\cite{cf}, except for seagull counter terms first
incorporated completely in Ref.\cite{texas} (see also Refs.\cite{tm,3nf}). We
emphasize that the former calculation treated the ``Born terms'' (in this case
the leading-order $\pi$-$N$ coupling and the Weinberg-Tomozawa $\pi \pi$-$N$
coupling) to nonstatic (subleading) order, but important (dominant) terms in the
chiral expansion were not treated until Ref.\cite{texas}. A review of
one-pion-exchange contributions to the Hamiltonian and to the charge form factor
(from electron scattering) was presented in the Appendix to \cite{cf}.  In
addition, all necessary formulae were developed for determining $\Delta
V_{2\pi}$, but the final integrals and spin-isospin commutators were not
performed.  This will be completed below.

The work of Refs.\cite{mec,cf} is based on a relativistic Lagrangian for
free nucleons plus a PV $\pi$-$N$ interaction plus a Weinberg-Tomozawa\cite{wt} 
$\pi \pi$-$N$ interaction. A Foldy-Wouthuysen reduction 
of this Lagrangian is made to the appropriate order in $1/M$ and this reduced
set is used in time-dependent perturbation theory in the usual way. It yields
potentials and pion-exchange-current operators. Because time-dependent 
perturbation theory leads to a variant of (time-independent) Schr\"odinger 
perturbation theory, the resulting potentials are energy-dependent. A mapping
technique [or (alternatively) perturbation theory manipulations] convert those
forms to an energy-independent potential to be used in Eq.~(1), with the 
kinetic energy given by Eq.~(4).

For specificity we list next the required Lagrangian terms,
${\mathcal{L}}^{(\Delta)}$, where $\Delta = 0, 1, \cdots$ represents powers of
$1/\Lambda$ (i.e., $\Lambda^{-\Delta}$) contained implicitly and explicitly in
the coefficients of the products of pion and nucleon fields.  The term
${\mathcal{L}}^{(3)}$ is only formally required to produce the necessary part of
the free relativistic nucleon energy, $T$, listed in Eq.~(4); it was not
required to calculate the $2\pi$-exchange force in Ref.\cite{cf}. The
${\mathcal{L}}^{(2)}$ terms (see Eq.~(6f) of Ref.\cite{mec}) are required only
to calculate the $\Delta V_{\pi}$ potential and are not needed for $\Delta
V_{2\pi}$.  Only ${\mathcal{L}}^{(0)}$ and ${\mathcal{L}}^{(1)}$ (see Eq.~(5) of
Ref.\cite{cf}) are required in order to produce $\Delta V_{2\pi}$. We have

$$
 {\cal L}^{(0)}  = \frac{1}{2}[\dot{\boldpi}^{2}-(\vec{\nabla}\boldpi)^{2}
          -m_{\pi}^{2}\boldpi^{2}] 
   + N^{\dagger}[i\partial_{0}-\frac{1}{4 f_{\pi}^{2}} \boldtau \cdot
         (\boldpi\times\dot{\boldpi})]N +\frac{g_{A}}{2 f_{\pi}} 
 N^{\dagger}\vecsig \cdot\vec{\nabla}(\boldtau\cdot\boldpi)N \, , \eqno (8)
$$
$$\eqalignno{
 &{\cal L}^{(1)} 
        =\frac{1}{2 M}\left [ - N^{\dagger}\vp^{\; 2}N
   -\frac{1+g_A^2\, (\mu - 1)}{4 f_{\pi}^{2}}N^{\dagger}\{\vp\, , \cdot
   (\boldtau \cdot \boldpi\times\vec{\nabla}\boldpi)\, \}N   \right.   & (9)\cr
& \left. +\frac{g_{A}\, (\mu + 1)}{4 f_{\pi}}
      N^{\dagger}\{ \vecsig \cdot \vp\, , \boldtau\cdot\dot{\boldpi} \} N 
      + \frac{g_A^2\, (\mu - 1)}{4 \fpi^2} 
        N^{\dagger}\boldpi \cdot \vec{\nabla}^2  \boldpi N\,
        -\frac{g_A^2\, (\mu - 1)}{8 \fpi^2} 
        N^{\dagger} \{ \vecsig \times \vp \, , \cdot
        \vec{\nabla} \boldpi^2  \} N \right ]  &\cr
  & +\frac{1}{f_{\pi}^{2}}N^{\dagger}[( -2c_1 m_{\pi}^{2} \boldpi^{2} -
        c_3 (\vec{\nabla}\boldpi)^{2} -
        \frac{1}{2} (c_4 + \frac{1}{4 M}) 
        \varepsilon_{ijk} \varepsilon_{abc} \sigma_{k} \tau_{c} 
        \partial_{i}\pi_{a}\partial_{j}\pi_{b}]N \, ,    & \cr}
$$
$$
 {\cal L}^{(2)}  = -\frac{g_{A}}{8 f_{\pi} M^2} 
 N^{\dagger}\{\vp^{\, 2} , \vecsig \cdot \vec{\nabla}
 (\boldtau\cdot\boldpi) \}N + 
 \frac{g_A\, \mu}{16 \fpi M^2} N^{\dagger} \{ \vecsig \cdot \vp \, , 
 \{\vp \, , \cdot \vec{\nabla} (\boldtau\cdot\boldpi) \} \} N \, , \eqno (10)
$$
$$
 {\cal L}^{(3)} =\frac{1}{8 M^3} N^{\dagger}\vp^{\; 4}N \, , \eqno (11)
$$

Terms not required in what follows have not been listed, and this includes all
short-range $NN$ operators.  Note that both terms in ${\mathcal{L}}^{(2)}$
depend on the average nucleon momentum $\vp$ and are constrained by the
requirements of relativity (as is ${\mathcal{L}}^{(3)}$ and four of the first
five terms of ${\mathcal{L}}^{(1)}$). The (nucleon) momentum-dependent terms
were developed by noting that the nucleon parts of ${\mathcal{L}}^{(0)}$ can be
represented by a (covariant) free-nucleon term plus a PV $\pi$-$N$ vertex
supplemented by a Weinberg-Tomozawa $\pi \pi$-$N$ interaction. Performing a 
Foldy-Wouthuysen
reduction of this set\cite{mec,cf} freezes out the antinucleon degrees of
freedom and leads to all terms explicitly proportional to inverse powers of $M$.
We will ignore all pion-momentum-dependent (form-factor) modifications of the
PV vertex. Although the coefficient of $\vp^{\; 2}$ in Eq.~(10) is completely
specified, this is solely due to our knowledge that the pion is a pseudoscalar
object.  Scalar- and vector-meson exchanges produce similar terms of opposite
sign to each other. Note the dependence on the off-shell parameter $\mu$, which 
arises from a redefinition of nucleon fields.  The resulting Lagrangian still 
satisfies chiral symmetry, but not on a term-by-term basis unless $\mu = 1$.

Results for OPEP appropriate for Eq.~(5) have been presented previously in many 
calculations and are summarized by Eq.~(A12) of Ref.\cite{cf}. We restrict 
ourselves to the two-nucleon sector; the three-nucleon sector is discussed 
extensively in that reference.  For continuity, we also revert to a convenient 
but old-fashioned notation for the effective $\pi$-$N$ coupling constant, $f$,
$$
f = \frac{G}{2 M} \sim \frac{g_A}{2\fpi}\,  , \eqno(12)
$$
where the second relation holds only if the Goldberger-Treiman\cite{gt} 
relation is exact. We find that to order ($1/\Lambda^3$)
$$
V_\pi = V^0_\pi - \frac{1}{2M^2} \{\vp^{\; 2} , V_\pi^0\} + 
\left(\frac{1}{2} - \nu \right) i [T_{NR}, U_G] \ + \ (\mu - 2\nu + 1) \, 
i [T_{NR}, U_E] \, , \eqno(13)
$$
where
$$
V_\pi^0 = f_0^2\, \mpi \, \boldtau_1 \cdot  \boldtau_2 \, 
\vecsig_1 \cdot \vec{\nabla}\, \vecsig_2 \cdot \vec{\nabla} \, h_0 (x)  
\, , \eqno(14) 
$$
$$
f_0^2 = \frac{f^2\, \mpi^2}{4 \pi} \, , \eqno(15)
$$
$$
\vx = \mpi (\vx_1 - \vx_2) \, , \eqno(16a)
$$
$$
h_0 = \frac{e^{-x}}{x}\,  . \eqno(16b)
$$
Introduction of form factors into $h_0$ is straightforward, but for simplicity
will not be done here.  Our primary interest is the tail of this force, as
analyzed by the Nijmegen PWA\cite{nijm}. We also require the Gross ($U_G$) and
equivalence ($U_E$) unitary transformations\cite{cf}
$$
U_G = \frac{1}{4M} \, \{\vp \, \cdot ,\vx \; V_\pi^0 (\vx)/\mpi\}\,  , \eqno(17)
$$
and
$$
U_E = \frac{f_0^2}{8 M}\, \boldtau_1 \cdot \boldtau_2 \, 
\left( \{ \vecsig_1 \cdot \vp , 
\vecsig_2 \cdot \vec{\nabla}\, h_0(x) \} + 
\{\vecsig_2 \cdot \vp, 
\vecsig_1 \cdot \vec{\nabla} \, h_0 (x)\} \right) \,  , \eqno(18)
$$
where here and elsewhere $\vec{\nabla}$ signifies $\vec{\nabla}_x$.

Although the second term in $V_\pi$ is simple in form (and represents the
expansion of $\frac{M}{E_0} V^0_\pi \frac{M}{E_0}$, where $E_0 = \sqrt{\vp^{\;
2} + m^2}$, in the CM frame), the other terms are much more complicated and
involve the coupling of (nucleon) spin and momentum.  This type of tensor
coupling is the origin of the lower than usual $P_D$ of the deuteron for the
traditional Bonn potential models, and subsequently for a higher triton binding
energy than other models\cite{da}.  Clearly, the last two terms in Eq.~(13) can
be neglected on-shell, or by {\bf choosing} $\nu = 1/2$ and $\mu = 0$ they can 
be neglected off-shell, as well.  This is a popular choice.  Note that the two 
factors of $(\frac{M}{E_0})$ arise from our nucleon normalization.  Rather than 
use the {\it covariant} normalization convention (current matrix elements are 
manifestly covariant and nucleon normalization factors are kept separate in 
$S$-matrix elements or are incorporated into covariant phase-space factors) we 
revert to the {\it invariant} convention (normalization factors are incorporated
into fields and the total charge, for example, is a Lorentz invariant) that is
consistent with a Foldy-Wouthuysen reduction and is conventional in nuclear
problems.

We further note that Sugawara and Okubo encountered both $U_G$ and $U_E$ in the
course of their derivation of the one- and two-pion-exchange potentials. Their
conversion of intermediate results to $V_\pi^0 - \{\vp^{\; 2} , V_\pi^0\}/2M^2$
is equivalent to choosing $\nu = 1/2$ and $\mu = 0$ in our result.  That choice 
has been recently called the ``minimal nonlocality'' [MNL] choice\cite{atom}, 
because it eliminates the complicated $\vecsig \cdot \vp$ terms that otherwise 
arise. This type of complexity matters little for momentum-space computational
approaches, but is a significant impediment to configuration-space calculations,
which are now the dominant approach for high-accuracy computations of $A >
3$ \cite{joe}.  Minimal nonlocality is a significant simplification.

As stated earlier, all but the final integrals and spin-isospin manipulations
for obtaining the nonstatic $\Delta V_{2\pi}$ were performed in Ref.\cite{cf} 
during the calculation of the nonstatic ``Born-term'' contributions to the
$2\pi$-exchange three-nucleon force.  We require Eqs.~(15) and (16) (direct-
plus crossed-box diagrams), Eq.~(18a) (``uncrossed'' or overlapping time-ordered
diagrams), Eq.~(20) (seagulls with no time derivatives), and Eq.~(22a) (the
Weinberg-Tomozawa seagull) of that reference. It was necessary to subtract the
iterated contribution of OPEP, $V_\pi$, since this is automatically included in
the solution of the RSE, Eq.~(5).  This subtraction led to an energy-dependent
potential that was mapped into an energy-independent form (this can be achieved
by many different equivalent techniques\cite{evgeni}).  Unitary equivalences of 
$2\pi$-range (such as in Eqs.~(18a) and (22a) of Ref.\cite{cf}) were eliminated 
by converting them to $3\pi$-range (which terms we are consistently ignoring). 
By choosing to develop an energy-independent potential we have eliminated from 
consideration the first of the three ambiguities discussed in the introduction.

The resulting nonstatic $2\pi$-exchange nucleon-nucleon force is given in its
most general form appropriate for $\mu$ and $\nu$ off-shell ambiguities by

$$
\Delta V_{2\pi}^{\mu , \nu} = \frac{f_0^4 \mpi^2}{2M} \  \left \{\vec{L} \cdot 
\vec{S} \left[3 \left(8g^2 - 2x g g^\prime (\mu - 2\nu - 3)\right) - 2 
\boldtau_1 \cdot \boldtau_2 \left(8g^2 (\frac{1}{g_A^2} - 1) - 
2x g g^\prime (\mu - 2\nu + 1)\right) \right] \right.
$$
$$
+  \biggl [ 3\left( - x g (4g^\prime + x g^{\prime\prime}) (\mu - 3\nu) 
- (1 - \nu) 
x^2g^\prime (2g^\prime + x g^{\prime\prime}) + 16 \overline{c}_3 (3g^2 + 
x g^\prime(2g +x g^\prime)) + 32 \overline{c}_1 x^2g^2 \right) 
$$
$$
 - 2 \boldtau_1 \cdot \boldtau_2\, \left(-x g(4g^\prime + x g^{\prime\prime}) 
(\mu - 3\nu - 2) + (1 + \nu) x^2 g^\prime(2g^\prime + x g^{\prime\prime}) + 
\frac{2}{g_A^2} (3g^2 + x g^\prime(2g + x g^\prime))\right) \biggr ] 
$$
$$
+ \frac{1}{3}(3(1 + \nu) + 2 (1 - \nu) \boldtau_1 \cdot \boldtau_2) 
\left[ S_{12} (x g g^\prime + x^2(g^{\prime\, 2} + g g^{\prime\prime})) - 2 
\vecsig_1 \cdot \vecsig_2 (4 x g g^{\prime} + x^2(g^{\prime \, 2} + 
g g^{\prime\prime})) \right]
$$
$$
\left. + \frac{8}{3} \boldtau_1 \cdot \boldtau_2 \left( 4 
\overline{c}_4 + \frac{1}{g^2_A} \right) \left[ S_{12} (x g g^\prime) - 
\vecsig_1 \cdot \vecsig_2 (3g^2 + 2x g g^\prime) \right] \right \} \, ,
\eqno(19)
$$
where 
$$
g(x) = \frac{1}{x} \ \frac{d\, h_0(x) }{d x \, \ \ \ \ \ }  
= -e^{-x} (\frac {1}{x^2} + 
\frac {1}{x^3})\,  , \eqno(20)
$$
and 
$$
\overline{c}_i = c_i M/g_A^2  \, .\eqno(21)
$$
Each type of force in Eq.~(19) (spin-orbit, central, tensor, and spin-spin) 
depends on the off-shell parameters. For comparison purposes (see Eq.~(7)) we 
also find
$$
\frac{V_\pi^2}{4M} = \frac{f_0^4 \mpi^2}{12 M} (3 - 2 \boldtau_1 \cdot 
\boldtau_2) \left[ 2 S_{12}\, (x g g^\prime) - 2 \vecsig_1 \cdot \vecsig_2 
(3g^2 + 2x g g^\prime) + 3 (3g^2 + 2x g g^\prime + x^2 g^{\prime \, 2}) \right]
\, . \eqno(22)
$$

Note that only two types of isospin operator are possible:  direct type $(3 - 2
\boldtau_1 \cdot \boldtau_2)$ and crossed type $(3 + 2 \boldtau_1 \cdot
\boldtau_2)$. Both uncrossed diagrams (with slanted time-ordered pion
propagators) and reducible diagrams (with no pions at some intermediate time)
are of direct type.  This separation facilitates comparisons between different 
calculations of certain diagrams.

The crucial elements in this calculation are the chiral seagulls.  These were
first calculated in Ref.\cite{texas}.  Written in the manner of Eq.~(21), the
$\overline{c}_i$ should be dimensionless numbers whose magnitudes are $\sim 1$ 
\cite{ndpc}. Because the low-mass $\Delta$ isobar plays such a large role in 
$c_3$ and $c_4$, these values are even larger and they play a dominant role in 
$\Delta V_{2 \pi}$.  The remaining terms depend only on $f$.  They were first 
calculated in an energy-independent form by Sugawara and Okubo.  Their 
calculation suppressed negative-energy states, which contribute to our $\pi$-$N$
seagull amplitudes. They noted, however, that their PV calculation 
(corresponding to $\mu = 1$ and $\nu = 0$) had very weak pair terms, and indeed 
${\mathcal{L}}^{(0)}$ and ${\mathcal{L}}^{(1)}$ have no seagulls if we eliminate
the $c_i$, the Weinberg-Tomozawa terms (independent of $g_A$), and set $\mu = 
1$. For this reason their calculation of the relativistic direct- and 
crossed-box diagrams for PV coupling was complete.

We have verified their results for individual terms both before and after the
unitary transformations they performed.  Their final result corresponds to $\mu
= 0$, $\nu = 1/2$ in Eq.~(19) with the
aforementioned seagull terms dropped. Thus, the S-O PV calculation is the first
correct calculation of the subtracted box and crossed-box graphs. Their PS
calculation (corresponding to $\mu = -1$, $\nu = 0$) is however missing some
seagull terms, although we have verified all of their other calculated
contributions. Although S-O unnecessarily approximated $\sqrt{p^2 + M^2} - M$ by
$p^2/2M$ in the free-nucleon energy, it is clear from their calculation that
$V_\pi + V^0_{2\pi} + \Delta V_{2\pi}^{MNL}$ is to be substituted for $V$ in
Eq.~(5).

\section{Specific Forms of $\Delta V_{2\pi}$}

Having developed a new and general form for $\Delta V_{2\pi}$ corresponding to 
the $\mu$ and $\nu$ off-shell ambiguities, we now relate this form to specific
cases of potentials and procedures that are in common use today. For the special
(and preferred) MNL case ($\mu = 0$, $\nu = 1/2$) one finds
$$
\Delta V^{MNL}_{2\pi} = \frac{f_0^4 \mpi^2}{4M} \ \Biggl \{16 \vec{L} \cdot 
\vec{S} \left [ 3(g^2 + x g g^\prime) - 2 \boldtau_1 \cdot \boldtau_2 \,
g^2 \left(\frac{1}{g^2_A} - 1 \right) \right ] 
$$
$$
+ \Biggl [ 3 \left (3 x g (4g^\prime + x g^{\prime\prime}) - x^2g^\prime 
(2g^\prime + x g^{\prime\prime}) + 32 \overline{c}_3 (3g^2 + x g^\prime (2g + 
x g^\prime)) + 64 \overline{c}_1 x^2 g^2 \right)
$$
$$
- 2 \boldtau_1 \cdot \boldtau_2 \left( 7 x g (4g^\prime + x g^{\prime\prime}) 
+ 3 x^2g^\prime (2g^\prime + x g^{\prime\prime}) + \frac{4}{g^2_A} (3g^2 + 
x g^\prime (2g + x g^\prime))\right) \Biggr ]
$$
$$
+\frac{1}{3} (9 + 2 \boldtau_1 \cdot \boldtau_2) \left [S_{12} (x g g^\prime + 
x^2(g^{\prime \, 2} + g g^{\prime\prime})) - 2 \vecsig_1 \cdot \vecsig_2 
(4x g g^\prime + x^2(g^{\prime \, 2} + g g^{\prime\prime})) \right ]
$$
$$
+ \frac{16}{3} \boldtau_1 \cdot \boldtau_2 (4 \overline{c}_4 + 
1/g^2_A) \left [ S_{12} (x g g^\prime) - \vecsig_1 \cdot \vecsig_2 
(3g^2 + 2x g g^\prime) \right ] \Biggr \}  \, . \eqno (23)
$$

The form appropriate for the Nijmegen PWA (cf. Eq.~(7)) is then given by
$$
\Delta V_{2\pi}^{Nij} = \Delta V_{2\pi}^{MNL} + \frac{V_\pi^2}{4M} \, , 
\eqno(24)
$$
and should be used in Eq.~(6) with an OPEP in the form
$$
V^{Nij}_{\pi} = \left\{ \frac{E_0}{2M} , \frac{M}{E_0}V_{\pi}^0 (\vr) 
\frac{M}{E_0}\right\} \, , \eqno(25)
$$
where $E_0 = \sqrt{\vp^{\; 2} + M^2}$ is an operator. 

The usual approach of the Nijmegen group, however, is to approximate the 
operator-valued factors of $E_0$ in Eq.~(25) by their on-shell value: 
$E_{\rm os} = \sqrt{k^2 + M^2}$. This changes their OPEP to
$$
\overline{V}_\pi^{Nij} = \frac{M}{E_{\rm os}} \ V_\pi^0  \, , \eqno(26)
$$
and the difference of $V^{Nij}_{\pi}$ and $\overline{V}^{Nij}_{\pi}$ modifies
the required nonstatic $2 \pi$-exchange potential to (second-order perturbation 
theory is the easiest way to see this\cite{note})
$$
\Delta \overline{V}_{2\pi}^{Nij} = \Delta V_{2\pi}^{MNL} + \frac{3\, 
V_\pi^2}{4M}  \, . \eqno(27)
$$
That is, changing from operator-valued kinematical factors to on-shell
(c-number) ones merely adds $\frac{2\, V_\pi^2}{4M}$ to the ``form'' correction
factor displayed in Eq.~(24). 

In a similar fashion we note that the new Urbana\cite{jun} relativistic 
potential model corresponds to Eq.~(5) with
$$
V^{\rm Urb} = (\frac{M}{E_0} V_\pi^0 \frac{M}{E_0} + V_{2\pi}^0 + 
\Delta V_{2\pi}^{MNL}) + \cdots \, .  \eqno(28)
$$
In addition, the Bonn potential\cite{bonn} incorporated into an equation with
relativistic kinematics requires a slightly different version of $\Delta 
V_{2 \pi}^{\mu,\nu}$ corresponding to $\mu=-1$ and $\nu=\haf$ in Eq.~(19): 
$$
\Delta V_{2\pi}^{\rm Bonn} = \Delta V_{2\pi}^{-1, 1/2} \, . \eqno(29)
$$

Equations~(23-29), as well as Eq.~(19), are our principal results. Specifying 
the form of the equation to be solved and the form of OPEP eliminates the three 
types of ambiguity discussed in the introduction.

Finally, a very interesting and unusual calculation (from the nuclear physics
perspective) was recently performed\cite{munich}.  The one-pion-exchange
potential was calculated on-shell, and the (wave-function-normalization) factor
$(M/E_0)^2$ was extracted as a kinematical (flux) factor.  Thus, the OPEP was
chosen to be the nonrelativistic $V_\pi^0$.  Subtracting ($M / E_0$) times the
iterated OPEP (using a nonrelativistic Green's function) they found a $\Delta
V_{2\pi}$ in the form
$$
\Delta V_{2\pi} ({\rm Ref.}\cite{munich}) = \Delta V_{2\pi}^{MNL} + 
\frac{3 V_\pi^2}{4M} \, , \eqno (30)
$$
and did not relate this result to any particular dynamical equation.  We 
have verified that Eq.~(30) corresponds to the conditions of Ref.\cite{munich}.

\section{Summary}

In summary, we have completed an old-fashioned calculation of the nonstatic
two-pion-exchange nucleon-nucleon potential that emphasizes and highlights the
off-shell nature of OPEP.  Different versions were developed that correspond to
different $(\mu, \nu)$ off-shell parameters, and to using relativistic or
nonrelativistic kinetic energies in the Schr\"{o}dinger Equation, or using
operator-valued or on-shell kinematic factors. These new results would
correspond to the Nijmegen\cite{psa}, Urbana\cite{jun}, and Bonn\cite{bonn}
approaches, and further manipulation leads to the Munich result\cite{munich}. 
These various approaches in perturbation theory correspond to different
subtractions in the second-order iteration of OPEP and represent different
off-shell versions of that potential.
 
\pagebreak

\noindent {\bf Acknowledgements}

This work was performed under the auspices of the United States Department of 
Energy. I have benefited greatly from many conversations with R.\ Timmermans 
(KVI) and J.\ J.\ de Swart (Nijmegen), and from a very helpful exchange with 
U.\ van Kolck.


\begin{thebibliography}{99}

\bibitem{tm} S.\ A.\ Coon, M.\ D.\ Scadron, P.\ C.\ McNamee, B.\ R.\ Barrett,
D.\ W.\ E.\ Blatt, and B.\ H.\ J.\ McKellar, {\it Nucl.\ Phys.\ }{\bf A317}, 
242 (1979). The chiral seagulls ($\sim c_i$) were first encountered here in 
a calculation of three-nucleon forces. The formalism used current algebra 
rather than Lagrangians.

\bibitem{3nf} J.\ L.\ Friar, D.\ H\"uber, and  U.\ van Kolck, 
{\it Phys.\ Rev.\ C} {\bf 59}, 53 (1999). The three-nucleon forces from the
chiral seagulls of Ref.\cite{texas} are rederived here in the context of 
another type of off-shell behavior arising from field redefinitions. These
terms should be added to the ``Born-term'' results of Ref.\cite{cf} to obtain
complete two-pion-exchange three-nucleon forces of subleading order 
(corresponding to an energy-independent OPEP).

\bibitem{quasi} J.L. Friar, {\it Phys. Rev. C} {\bf 22}, 796 (1980). A great 
number of quasipotential equations exist and several of these are treated in
this reference. Others not treated are cited. Some interesting and more recent 
treatments are: D.\ R.\ Phillips and S.\ J.\ Wallace, {\it Phys.\ Rev.\ C} {\bf
54}, 507 (1996); A.\ Stadler and F.\ Gross, {\it Phys.\ Rev.\ Lett.} {\bf 78},
26 (1997); V.\ Pascalutsa and J.\ A.\ Tjon, {\it Phys.\ Lett.} {\bf B435}, 245
(1998).

\bibitem{mec} J.\ L.\ Friar, {\it Ann.\ Phys.} (N.Y.) {\bf 104}, 380 (1977); 
this is the seminal article for our approach. Energy-dependent terms in the
pion-exchange potential were noted here (and removed) as part of a 
pion-exchange-current calculation. J.\ Adam, Jr., H.\ G\"oller, and H.\ 
Arenh\"ovel, {\it Phys.\ Rev.\ C} {\bf 48}, 370 (1993) is a modern version of 
this work containing numerical results.

\bibitem{cf}  S.\ A.\ Coon and J.\ L.\ Friar, {\it Phys.\ Rev.\ C} {\bf 34}, 
1060 (1986). Eq.~(33b) should equal Eq.~(27b), but contains a typographical 
error: the indices $i$ and $j$ should be interchanged in Eq.~(33b) [J. Adam,
Private Communication]. Consistency of potential models under Lorentz boosts
was discussed in this work, as well as in Refs.\cite{mec,quasi}. The dependence
of our $2\pi$-exchange interaction on the overall nuclear momentum (not listed 
in the present paper, but arising from the spin-orbit interaction) is 
constrained by Eqs.~(36a) and (A21) of Ref.\cite{cf}, and can be shown to be 
satisfied.

\bibitem{sl} F.\ B.\ Hildebrand, {\it Methods of Applied Mathematics},
(Prentice-Hall, Englewood Cliffs, 1952).

\bibitem{bw} K.\ A.\ Brueckner and K.\ M.\ Watson, {\it Phys.\ Rev.} {\bf 92},
1023 (1953).

\bibitem{tmo} M. Taketani, S. Machida, and S. Ohnuma, {\it Prog. Theor. Phys.
(Kyoto)} {\bf 7}, 45 (1952). There are typographical errors in TMO that are
corrected in, S.\ Machida, {\it Prog.\ Theor.\ Phys.\ (Suppl.)} {\bf 39}, 91
(1967).

\bibitem{fc} J.L. Friar and S.A. Coon, {\it Phys. Rev. C} {\bf 49}, 1272 
(1994).

\bibitem{texas}  C.\ Ord\'o\~nez and U.\ van Kolck, {\it Phys.\ Lett.} 
{\bf B291}, 459 (1992); C.\ Ord\'o\~nez, L.\ Ray, and U.\ van Kolck, {\it
Phys.\ Rev.\ Lett.} {\bf 72}, 1982 (1994); U. van Kolck, {\it Phys. Rev. C} {\bf
49}, 2932 (1994); C.\ Ord\'o\~nez, L.\ Ray, and U.\ van Kolck, {\it
Phys.\ Rev.\ C} {\bf 53}, 2086 (1996).

\bibitem{s-o}  M.\ Sugawara and S.\ Okubo, {\it Phys.\ Rev.} {\bf 117}, 
605 and 611 (1960). The first reference treats PS coupling and the second PV.

\bibitem{rel} J.\ L.\ Forest, V.\ R.\ Pandharipande, and J.\ L.\ Friar, {\it 
Phys. Rev. C} {\bf 52}, 568 (1995).

\bibitem{min} J.\ L.\ Friar, in {\it Mesons in Nuclei}, ed.\ by M.\ Rho and 
D.\ Wilkinson, (North-Holland, Amsterdam, 1979), p.\ 597. A simple example of
how to eliminate energy dependence is worked out two ways, one using algebraic 
methods identical to those used in \cite{cf} and the second using unitary 
transformations (the Foldy-Wouthuysen transformation).

\bibitem{bonn} R.\ Machleidt, K.\ Holinde, and C.\ Elster, {\em Phys.\ Rep.}
{\bf 149}, 1 (1987). The Bonn potentials correspond to $\mu= -1$. This
generates nonlocality that results in a lower deuteron D-state percentage and
a higher triton binding energy.

\bibitem{atom} J.\ L.\ Friar, J.\ Martorell, and D.\ W.\ L.\ Sprung, 
{\it Phys.\ Rev.\ A} {\bf 56}, 4579 (1997).

\bibitem{jun} J.\ L.\ Forest, V.\ R.\ Pandharipande, and A.\ Arriaga, (submitted
to Physical Review C); nucl-th/9805033.

\bibitem{cpt} S.\ Weinberg, {\it Physica} {\bf 96A}, 327 (1979); 
{\it Nucl.\ Phys.} {\bf B363}, 3 (1991); {\it Phys.\ Lett.} {\bf B251}, 288 
(1990); {\it Phys.\ Lett.} {\bf B295}, 114 (1992).

\bibitem{bkm} V.\ Bernard, N.\ Kaiser, and  U.-G.\ Mei{\ss}ner, {\it Int.\ J.\ 
Mod.\ Phys.\ }{\bf E4}, 193 (1995) is a comprehensive current review that is 
indispensable to the practitioner. 

\bibitem{bs} R.\ Blankenbecler and R.\ Sugar, {\it Phys.\ Rev.} {\bf 142}, 1051
(1966).

\bibitem{fritz} F.\ Coester, S.\ C.\ Pieper, and F.\ J.\ D.\ Serduke, {\it 
Phys.\ Rev.\ C} {\bf 11}, 1 (1975). This is the earliest use of the squaring
trick (of which we are aware) for relating the RSE and Schr\"odinger-like
equations. It was also briefly used in Ref.\cite{mec}.

\bibitem{psa} V.\ G.\ J.\ Stoks, R.\ A.\ M.\ Klomp, M.\ C.\ M.\ Rentmeester, 
and J.\ J.\ de Swart, {\it Phys. Rev. C} {\bf 48}, 792 (1993). The procedures 
of this reference allow $\overline{V}$ in Eq.~(7) to be fit to $N$-$N$ data. In
principle this equation should be inverted to obtain the $V$ required for 
Eq.~(6), but in practice this is not done.

\bibitem{rll} J.\ Carlson, V.\ R.\ Pandharipande, and R.\ Schiavilla,
{\it Phys.\ Rev.} C {\bf 47}, 484 (1993).

\bibitem{ndpc} A.\ Manohar and H.\ Georgi, {\it Nucl. Phys.} {\bf B234}, 189 
(1984); H.\ Georgi, {\it Phys. Lett.} {\bf B 298}, 187 (1993).

\bibitem{tom} Th.\ A.\ Rijken, {\it Ann.\ Phys.} {\bf 208}, 253 (1991). There
are too many calculations to list them all. We find this recent one 
particularly useful, and it contains many references to older work.

\bibitem{wt} S.\ Weinberg, {\it Phys.\ Rev.\ Lett.} {\bf 17}, 616 (1966);
Y.\ Tomozawa, {\it Nuovo Cimento A} {\bf 46}, 707 (1966).

\bibitem{gt} M.\ L.\ Goldberger and S.\ B.\ Treiman, {\it Phys.\ Rev.} {\bf
110}, 1178 (1958); Y.\ Nambu, {\it Phys.\ Rev.\ Lett.} {\bf 4}, 380 (1960).

\bibitem{nijm} M.\ C.\ M.\ Rentmeester, R.\ G.\ E.\ Timmermans, J.\ L.\ Friar, 
and J.\ J.\ de Swart, {\it Phys.\ Rev.\ Lett.} (submitted).

\bibitem{da} B.\ Desplanques and A.\ Amghar, {\it Z.\ Phys.\ A (Hadrons and 
Nuclei)} {\bf 344}, 191 (1992).

\bibitem{joe} J.\ Carlson and R.\ Schiavilla, {\it Rev. Mod. Phys.} {\bf 70},
743 (1998).

\bibitem{evgeni} E.\ Epelbaoum, W.\ Gl\"ockle, and Ulf-G.\ Mei\ss ner, 
{\it Nucl.\ Phys.} {\bf A637}, 107 (1998). This recent work discussed some of 
the problems associated with energy-dependent potentials.

\bibitem{note} An off-shell function of $\vp^{\; 2}/M^2$ becomes an on-shell
function of $k^2/m^2$. The difference of the two (expanded in powers of $1/M$)
is proportional to $(\vp^{\; 2} -k^2)/M^2$, or to the inverse nonrelativistic 
Green's function. In perturbation theory this generates an effective potential 
that equals a numerical factor times $V^2/M$. The numerical factor depends on 
the specific form of the function.

\bibitem{munich} N.\ Kaiser, R.\ Brockmann, and W.\ Weise, {\it Nucl.\ Phys.}
{\bf A625}, 758 (1997).

\end{thebibliography}
\end{document}